\documentclass[doublecol,linenumbers]{epl2} 

\usepackage{amsmath,amsfonts,amsthm,amssymb}
\usepackage[colorlinks=true,linkcolor=blue,urlcolor=blue,citecolor=blue]{hyperref}
\usepackage{epstopdf}

\title{Magnetic field--induced modification of selection rules for Rb D$_2$ line monitored by selective reflection from a vapor nanocell}
\shorttitle{Magnetic field induced modification} 

\author{E. Klinger\inst{1,2}, A. Sargsyan\inst{1}, A. Tonoyan\inst{1}, G. Hakhumyan\inst{1}, A. Papoyan\inst{1}, C. Leroy\inst{2,3} \& D. Sarkisyan\inst{1}}
\shortauthor{E. Klinger \etal}

\institute{                    
  \inst{1} Institute for Physical Research, NAS of Armenia, Ashtarak-2, 0203, Armenia\\
  \inst{2} Laboratoire Interdisciplinaire Carnot de Bourgogne, UMR CNRS 6303, Universit\'e Bourgogne - Franche-Comt\'e, BP 47870, 21078 Dijon Cedex, France\\
  \inst{3}{Institute of Physics and Technology, National Research Tomsk Polytechnic University, Tomsk 634050, Russia}
}
\pacs{Atomic physics, Intensities and shapes of atomic spectral lines, Zeeman and Stark effects.}{}

\abstract{
Magnetic field-induced giant modification of the probabilities of five transitions of $5S_{1/2}, F_g=2 \rightarrow 5P_{3/2}, F_e=4$ of $^{85}$Rb and three transitions of $5S_{1/2}, F_g=1 \rightarrow 5P_{3/2}, F_e=3$ of $^{87}$Rb forbidden by selection rules for zero magnetic field has been observed experimentally and described theoretically for the first time. For the case of excitation with circularly-polarized ($\sigma^+$) laser radiation, the probability of $F_g=2, ~m_F=-2 \rightarrow F_e=4, ~m_F=-1$ transition becomes the largest among the seventeen transitions of $^{85}$Rb $F_g=2 \rightarrow F_e=1,2,3,4$ group, and the probability of $F_g=1,~m_F=-1 \rightarrow F_e=3,~m_F=0$ transition becomes the largest among the nine transitions of $^{87}$Rb $F_g=1 \rightarrow F_e=0,1,2,3$ group, in a wide range of magnetic field 200 -- 1000 G. Complete frequency separation of individual Zeeman components was obtained by implementation of derivative selective reflection technique with a 300 nm-thick nanocell filled with Rb, allowing formation of narrow optical resonances. Possible applications are addressed. The theoretical model is perfectly consistent with the experimental results.}

\begin{document}

\maketitle

\section{Introduction}
Alkali atoms (Rb, Cs, K, Na) are widely used in atomic spectroscopy due to strong atomic transitions from the ground state with wavelengths in visible and near-infrared regions (600 -- 900 nm), where cw narrow-band smoothly tunable diode lasers are available. Application areas of rubidium atoms include laser cooling experiments, information storage, spectroscopy, magnetometry, laser frequency stabilization, etc.\cite{Budker_atomic_physics,Auzinsh_polarized_atoms}. That is why any new knowledge of the behavior of Rb atomic transitions, particularly, exposed to an external magnetic field, is of high importance.\\
\indent It is well known that in quite moderate magnetic field $B$ the splitting of atomic energy levels to Zeeman sublevels deviates from the linear behavior, and the atomic transition probabilities undergo significant changes \cite{Auzinsh_polarized_atoms,Tremblay_1990,Scotto_2015,Sargsyan_2008,Sargsyan_2015_EPL,Sargsyan_2014,Hakhumyan_2012}. The most simple technique to monitor and study such modification is laser spectroscopy of atoms contained in an atomic vapor cell. For $B$ up to $\cong 1000$ G, the Zeeman split hyperfine transitions remain overlapped because of Doppler broadening, and sub-Doppler techniques are required to spectrally resolve transition probabilities of individual components \cite{Tremblay_1990}. Coherent Doppler narrowing in a thin vapor cell has been attained in \cite{Briaudeau_1998}.\\
\indent As it was demonstrated recently \cite{Sargsyan_2016,Sargsyan_2017_JOSA}, strong line narrowing can be achieved in Derivative Selective Reflection (DSR) spectra using an atomic vapor cell of nearly half-wavelength thickness ($L \approx \lambda/2$, where $\lambda$ is the resonant wavelength of laser radiation). These studies have been done for the case of $D_1$ lines of Rb and Cs. Among the advantages of DSR technique is proportionality of the recorded signal to atomic transition probability. In addition, the DSR resonance linewidth is practically immune against 10\% deviation of the cell thickness. These benefits make it convenient to use the DSR-method for studies of closely-spaced individual atomic transition components in a magnetic field.\\
\indent In the present paper we demonstrate that the DSR linewidth for Rb $D_2$ line is $\sim$ 50 MHz FWHM (opposed to $\sim$~500 MHz Doppler absorption linewidth in ordinary cells), which allows frequency separation of individual Zeeman components of hyperfine transitions and studying their transition probabilities in an external magnetic field. The DSR technique was employed to study for the first time, both experimentally and theoretically, the dynamics of giant modification of transition probabilities of $5S_{1/2},~ F_g=2 \rightarrow 5P_{3/2},~ F_e=4$ and $5S_{1/2},~F_g=1 \rightarrow 5P_{3/2},~F_e=3$ transitions of Rb $D_2$ line induced by a magnetic field varied in a wide range (up to 1000 G). These transitions are forbidden at $B = 0$ according to $\Delta F = 0, \pm1$ selection rule for the total momentum of atom. To the best of our knowledge, there were only a few articles where such type of transitions have been studied quantitatively \cite{Tremblay_1990,Scotto_2015,Sargsyan_2014}. 

\section{Experimental}

A nanometric-thin cell (NC) filled with Rb has been used in our experiment, allowing to obtain sub-Doppler spectra, and thus resolving a large number of hyperfine and  transitions and their Zeeman components. The general design of NC is similar to that described in \cite{Hakhumyan_2012,Dutier_2003_EPL}. Compact oven was used to set the needed temperature regime of $110~^{\circ}$C, which corresponds to the number density of Rb atoms $N\cong 10^{13}~$cm$^{-3}$. Adjustment of needed vapor column thickness without variation of thermal conditions was attained by smooth vertical translation of the cell-and-oven assembly. \\

\begin{figure}
\centering
\includegraphics[scale=0.32]{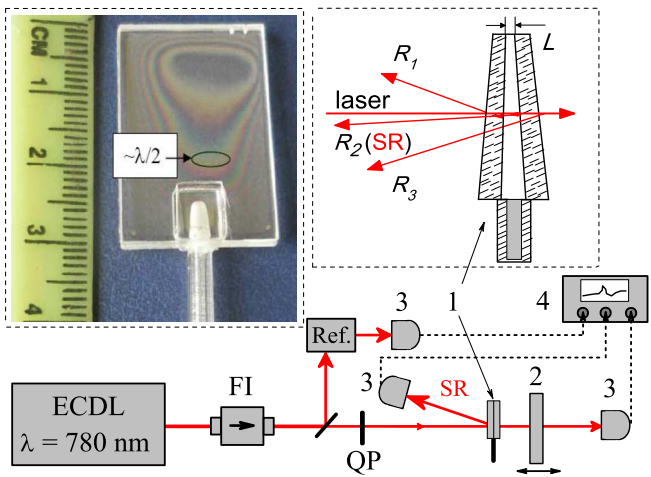}
\caption{Layout of the experimental setup: (ECDL) extended cavity diode laser; (FI) Faraday isolator; (1) nanocell filled with Rb inside the oven; (2) strong permanent magnet on a translation stage to produce $B$ $>$ 200 G field (for $B$ = 1 -- 200 G, the magnet was replaced by Helmholtz coils - not shown in the figure); (3) photodetector; (Ref.) Auxiliary unit for formation of frequency reference spectrum; (4) oscilloscope; (SR) selective reflection beam; (QP) quarter-wave plate. Upper left panel: photograph of the NC; interference fringes formed in the light reflected from the inner surfaces of the windows are seen; the oval marks the thickness region $300~\ab{nm}\leq L\sim\lambda/2\leq 400~\ab{nm}$. Upper right panel: the geometry of three beams reflected from the NC; the (SR) beam propagates along $R_2$.}
\label{fig-exp_setup}
\end{figure}

Schematic diagram of the optical part of experimental setup is shown in Fig.~\ref {fig-exp_setup}. A circularly polarized laser radiation beam ($\lambda =780~$nm, $P_L = 2~$mW, $\Delta\nu_{L} = 1~$MHz) resonant with Rb $D_2$ line was focused ($\varnothing= 0.5$~mm) at normal incidence angle onto a Rb NC with a vapor column thickness $L < \lambda /2 \approx 300~$nm. This optimum thickness was chosen to combine high spatial resolution, which is very important when using high-gradient field from permanent magnet, with acceptable broadening of the SR spectral linewidth (the latter increases with reduction of $L$). A calibrated strong permanent neodymium magnet placed near the rear window of the NC was used to produce strong longitudinal magnetic field controllable by changing the distance to the window. Noteworthy, extremely small thickness of the NC is advantageous for application of strong magnetic field produced by a permanent magnet: in spite of high field gradient, the variation of $B$-field inside the cell is several orders less than the applied value. To record DSR spectra, the laser radiation frequency was linearly scanned within up to 7 GHz spectral region covering the studied group of transitions. The nonlinearity of the scanned frequency ($\sim$ 1\% throughout the spectral range) was monitored by simultaneously recorded transmission spectra of a Fabry-P\'erot etalon (not shown). About 30\% of the pump power was branched to 3 cm-long Rb cell providing reference saturation absorption spectrum for $B = 0$. All the spectra were detected by photodiodes with amplifiers, and recorded by a four-channel digital storage oscilloscope Tektronix TDS 2014B.

\begin{figure*}
\centering
\includegraphics[scale=0.19]{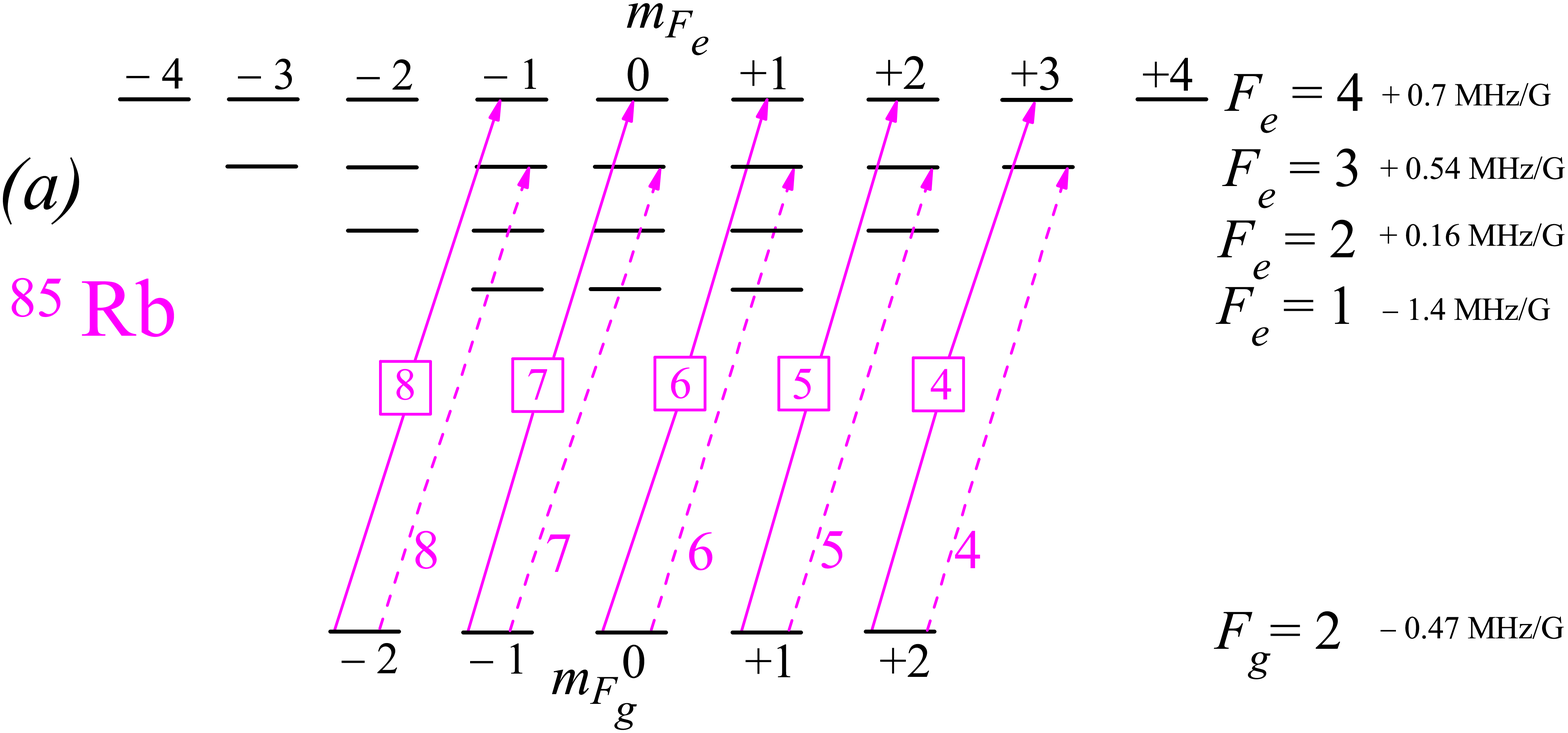}
\includegraphics[scale=0.19]{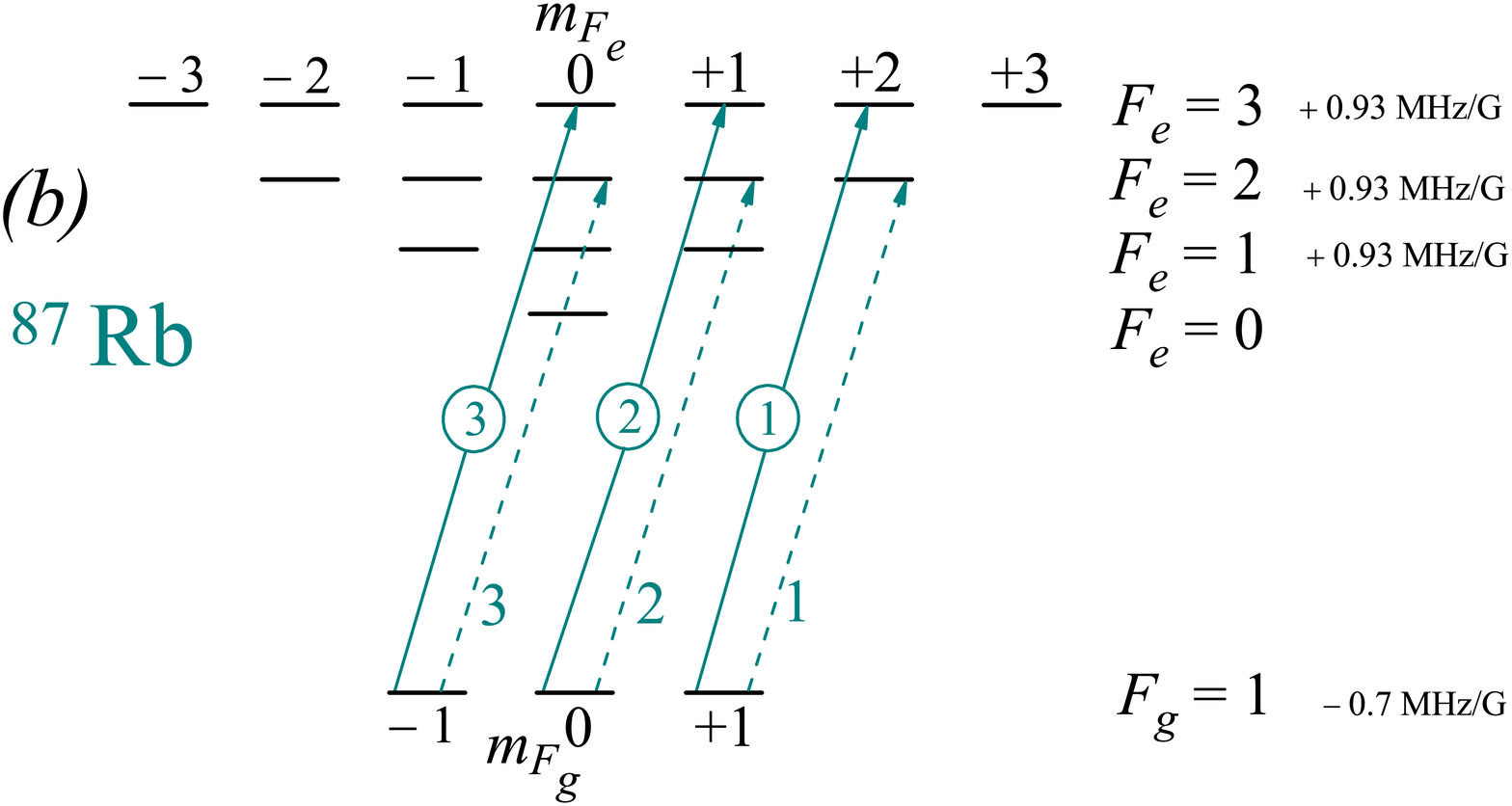}
\caption{Diagram of the relevant transitions between the Zeeman sublevels of Rb $D_2$ line with $\sigma^+$ (left-circular) laser excitation for the case of (a) $^{85}$Rb (nuclear spin $I = 5/2$), and (b) $^{87}$Rb (nuclear spin $I = 3/2$). Each transition is labeled to facilitate identification in the following graphs. Linear Zeeman shift rates are indicated next to each hyperfine level.}
\label{fig-atomic_diagram}
\end{figure*}

\begin{figure*}
\centering
\includegraphics[scale=0.63]{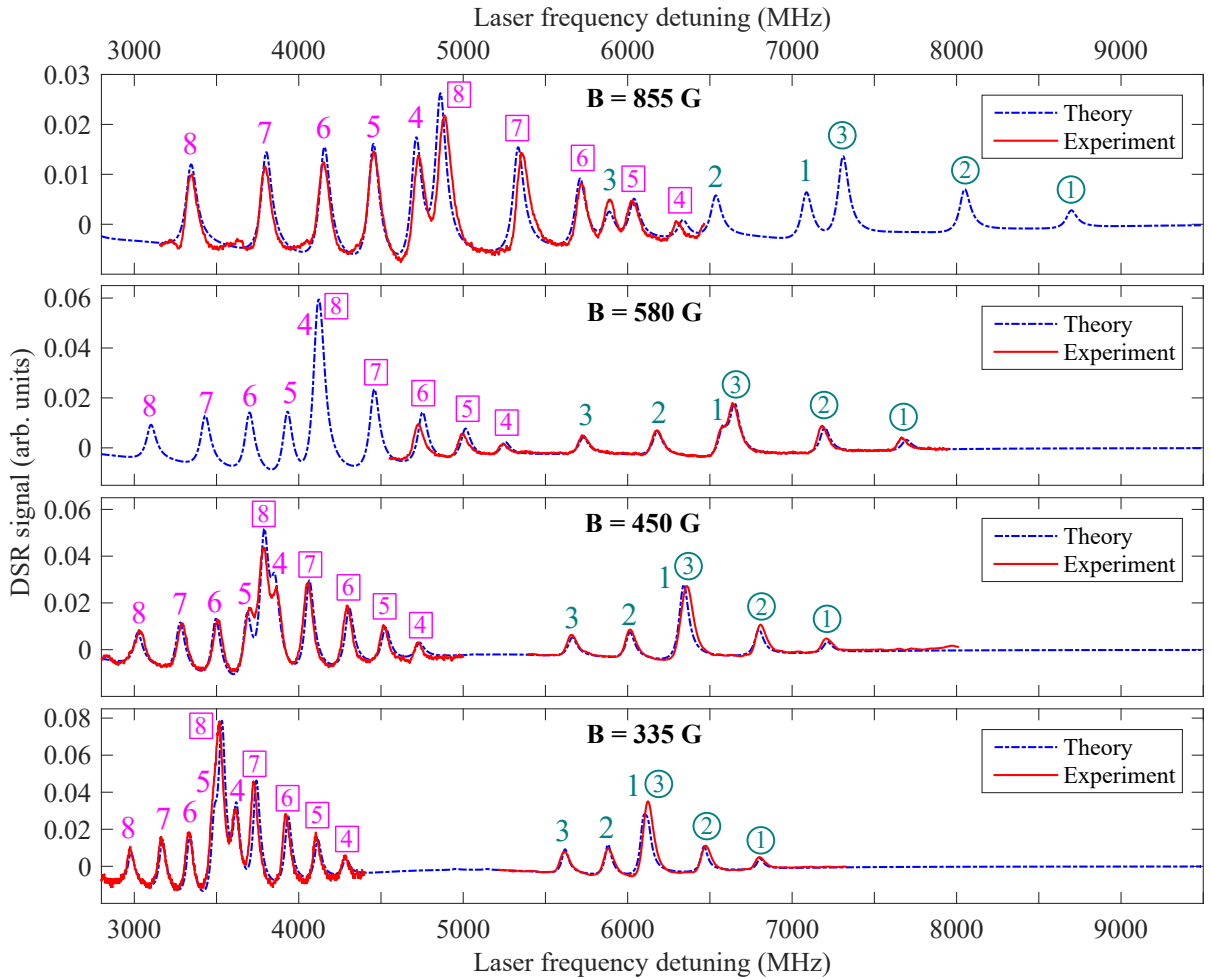}
\caption{Red solid lines: the DSR spectra for the Rb NC with the thickness $L \approx 300$~nm for $\sigma^+$ laser excitation ($P_L = 10~\mu$W) experimentally recorded at 4 values of the applied $B$-field: 335~G, 450~G, 580~G and 855~G. For labeling, see Fig.~\ref{fig-atomic_diagram}. Dashed blue lines: calculated spectra. Fragmentary presentation of experimental spectra is caused by mode hop--restricted continuity of the laser frequency scanning, verified by the Fabry-P\'erot cavity signal.}
\label{fig-theory_and_experiment}
\end{figure*}

\indent The diagram of relevant $\sigma^+$ components ($\Delta m_F = +1$) of $^{85}$Rb $D_2$ line transitions $F_g=2 \rightarrow F_e=3$ labeled 4 -- 8, and $F_g=2 \rightarrow F_e=4$ labeled $\fbox{4} - \fbox{8}$ are shown in Fig.~\ref{fig-atomic_diagram}(a). The transitions $\fbox{4} - \fbox{8}$ are forbidden for the zero magnetic field by selection rule $\Delta F = 0,\pm1$. The diagram of relevant $\sigma^+$ components ($\Delta m_F = +1$) of $^{87}$Rb $D_2$ line transitions $F_g=1 \rightarrow F_e=2$ labeled 1 -- 3, and $F_g=1 \rightarrow F_e=3$ labeled $\textcircled{1} - \textcircled{3}$ are shown in Fig.~\ref{fig-atomic_diagram}(b). The transitions $\textcircled{1} - \textcircled{3}$ are forbidden for zero magnetic field because of selection rule $\Delta F = 0,\pm1$. The transitions $F_g=2 \rightarrow F_e=1,2$ ($^{85}$Rb) and $F_g=1 \rightarrow F_e=0,1$ ($^{87}$Rb) are not shown in Fig.~\ref{fig-atomic_diagram}, since for $B > 300$~G their probabilities strongly reduce, making them practically undetectable in the DSR spectra. Note, that in the case of $\sigma^-$ excitation, the $F_g=2 \rightarrow F_e=4$ transitions of $^{85}$Rb and $F_g=1 \rightarrow F_e=3$ transitions of $^{87}$Rb have smaller DSR amplitudes than for the case of $\sigma^+$ for the same $B$  values (the reason of this distinction is out of scope of this paper, and will be presented elsewhere).\\
\indent The recorded DSR (real time derivative selective reflection) spectra of the Rb NC with thickness $L \approx 300$~nm for $\sigma^+$ laser excitation ($P_L = 10~\mu$W) and $B\approx335$~G, 450 G, 580 G and  855 G are shown in Fig.~\ref{fig-theory_and_experiment} (red solid lines). The dashed blue lines represent the results of theoretical modeling (see below). The sixteen transition components (for labeling, see Fig.~\ref{fig-atomic_diagram}) appear with $\approx$~60 MHz linewidth, being very well resolved. As it is seen, up to $B \approx$ 600~G the amplitudes of transitions $\fbox{7}$ and $\fbox{8}$ are the strongest among the transitions of $^{85}$Rb, $F_g=2 \rightarrow F_e=1,2,3,4$ (see also Fig.~\ref{fig-shift_and_proba} below).

\begin{figure*}
\centering
\includegraphics[scale=0.5]{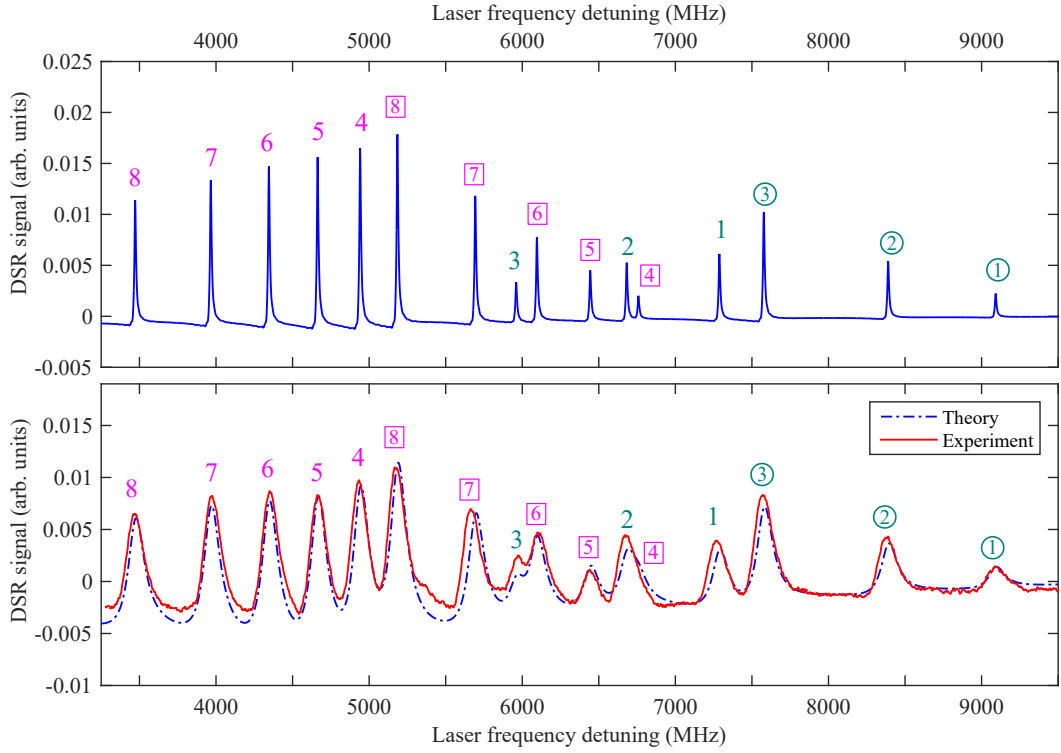}
\caption{Lower panel: experimentally recorded (red solid line) and calculated (dashed blue line) DSR spectra for the Rb NC with the thickness $L \approx 300$~nm for $\sigma^+$ laser excitation ($P_L = 10~\mu$W) and $B\approx 950$~G. Upper panel: theoretical DSR spectrum calculated with $\Gamma_{\ab{eff}} = 6$~MHz. For labeling, see Fig.~\ref{fig-atomic_diagram}.}
\label{fig-950G}
\end{figure*}

Similar DSR spectrum recorded for even higher value of magnetic field ($B\approx 950$~G) and otherwise invariable experimental conditions is presented by a red solid line in the lower panel of Fig.~\ref{fig-950G}, together with the modeled DSR spectrum (dashed blue line). All the sixteen labeled transition components are completely frequency resolved (except transitions 2 and $\fbox{4}$, which are overlapped). The upper panel shows the theoretical DSR spectrum calculated for effective broadening of $\Gamma_{\ab{eff}} = 6$~MHz corresponding to the natural transition linewidth, which is the minimum attainable width for the single laser beam experimental configuration.\\
\indent It is well known that the effect of applied field $B$ on the hyperfine structure is characterized by the parameter $B_0=A_{hfs}/\mu_B$, where $A_{hfs}$ is the hyperfine coupling constant for $5S_{1/2}$ and $\mu_B$ is the Bohr magneton \cite{Olsen_2011}. For $^{85}$Rb, $B_0 \approx 700$~G, and for $^{87}$Rb, $B_0 \approx 2400$~G. When $B_0 \ll B$ (such fields are treated as moderate), the splitting of energy levels is described by the total momentum of the atom $\boldsymbol{F} = \boldsymbol{J} + \boldsymbol{I}$ and its projection $m_F$, where $\boldsymbol{J} = \boldsymbol{L} + \boldsymbol{S}$ is the total momentum of the electron, and $\boldsymbol{I}$ is the spin momentum of the nucleus. When $B\gg B_0$, the coupling of $\boldsymbol{J}$ and  $\boldsymbol{I}$ ceases, and the splitting is described by the projections $m_J$ and $m_I$. The latter manifests the onset of hyperfine Paschen-Back (HPB) regime characterized by a number of peculiarities \cite{Olsen_2011,Weller_2012,Sargsyan_2015_OC}. Particularly, probabilities of the eight above mentioned transitions forbidden at $B$ = 0 reduce back to zero when $B\gg B_0$, so they are absent in the $m_J$ and $m_I$ basis. Note that the probabilities of $^{85}$Rb transitions labeled $\fbox{4} - \fbox{8}$ reduce faster with $B$ than the transitions $\textcircled{1} - \textcircled{3}$ of $^{87}$Rb (see Fig.~\ref{fig-shift_and_proba}). This is caused by the fact that $B_0$ value for $^{87}$Rb is larger than the one for $^{85}$Rb, consequently HPB regime for $^{87}$Rb occurs at larger magnetic fields. In total 20 atomic transitions remain in the complete HPB regime, 12 belonging to $^{85}$Rb, and 8 belonging to $^{87}$Rb \cite{Sargsyan_2015_OC}.

\section{Theoretical model and discussion}  
\begin{figure*}
\includegraphics[scale=0.80]{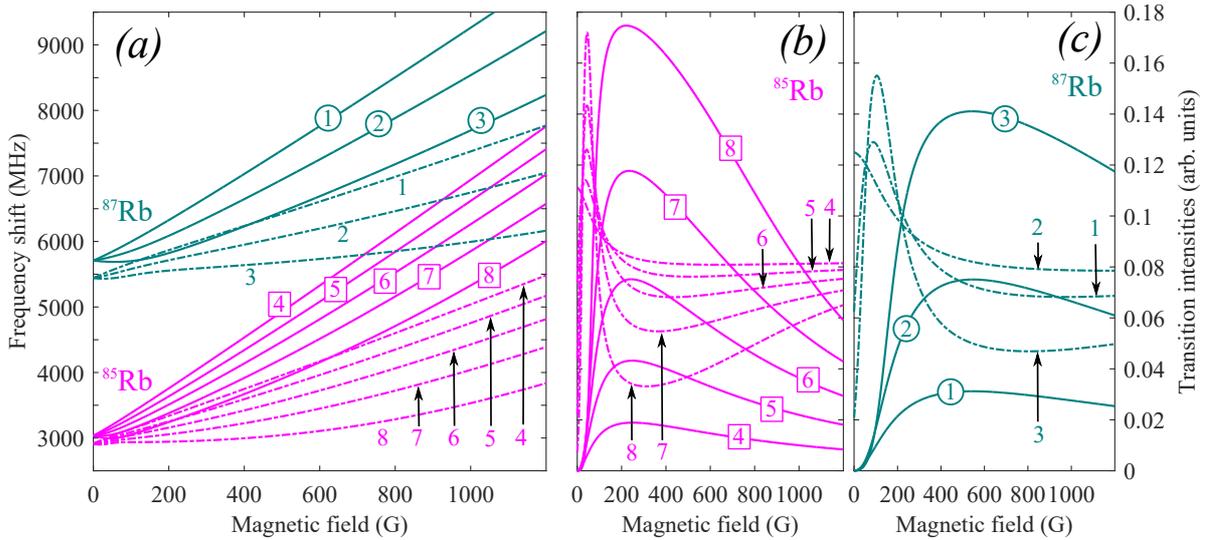}
\caption{(a) Magnetic field dependence of frequency shifts for Zeeman components of $F_g=2 \rightarrow F_e=3,4$ transitions of $^{85}$Rb $D_2$ line and $F_g=1 \rightarrow F_e=2,3$ transitions of $^{87}$Rb $D_2$ line for the case of $\sigma^+$ excitation. (b) Evolution of probabilities for Zeeman components of $F_g=2 \rightarrow F_e=3,4$ transitions ($^{85}$Rb) versus $B$-field for $\sigma^+$ excitation. (c) Evolution of probabilities for Zeeman components of $F_g=1 \rightarrow F_e=2,3$ transitions ($^{87}$Rb) versus $B$-field for $\sigma^+$ excitation. For labeling, see Fig.~\ref{fig-atomic_diagram}.}
\label{fig-shift_and_proba}
\end{figure*}

Below we summarize the main relations used to calculate the theoretical DSR spectra. The reader is invited to read the articles \cite{Dutier_2003_JOSA,Sargsyan_2017_JOSA} for more detailed information. Starting from the propagation equations of electric field inside the cell, one deduces the reflected signal given by
\begin{align}
\label{eq:Sr}
S_r & \approx  2 t_{10} E_{in} \Re\Big\lbrace r\big[1-\exp(-2ikL)\big] I_{SR}\Big\rbrace/|Q|^2~,
\end{align}
where $t_{10}$ and $r$ are transmission and reflection coefficients respectively, $Q=1-r^2\exp(2ikL)$ is the quality factor associated to the nanocell, and $I_{SR}$ is given in the linear regime of interaction by
\begin{align}
\label{eq:I_SR}
I_{SR}&=\big[1+r^2\exp(2ikL)\big]I^{lin}_{SR}-2r\exp(2ikL)I^{lin}_{T},
\end{align}
which depends on the induced polarization in the vapor through the integrals $I^{lin}_{SR}$ and $I^{lin}_{T}$, respectively expressed as
\begin{subequations}
\begin{align}
\label{eq:linear_I_SR}
I^{lin}_{SR}=C\int_{-\infty}^{+\infty}W(v)~h(\omega-\omega_{eg}, \Gamma,L,v)~dv \\
\label{eq:linear_I_T}
I^{lin}_{T}=C\int_{-\infty}^{+\infty}W(v)~g(\omega-\omega_{eg}, \Gamma,L,v)~dv
\end{align}
\end{subequations}
where $v$ is the speed of atoms in the cell, and $W(v)$ is the velocity distribution assumed to be Maxwellian, defined as $W(v)=(1/u\sqrt{\pi})\exp(-v^2/u^2)$ with $u$ the thermal velocity given by $u(T)=\sqrt{2k_BT/m}$ ($T$ is the vapor temperature, $k_B$ the Boltzmann's constant, and $m$ the atomic mass). The expressions for $h$ and $g$ are given in \cite{Dutier_2003_JOSA}, being the function of transition frequency $\omega_{eg}$, the cell thickness $L$, and transition width $\Gamma$, which includes the natural $\Gamma_{\text{nat}}$, collisional $\Gamma_{\text{col}}$, and inhomogenous $\Gamma_{\text{inhom}}$ broadening coefficients. In Eqs.~(\ref{eq:linear_I_SR}) and (\ref{eq:linear_I_T}), $C$ is a function of the atomic density $N$ and the dipole moment of transition from state $|g\rangle$ to state $|e\rangle$, and reads
\begin{align}
C= \frac{N t_{10}E_{in}}{4\hbar\epsilon_0Q}|\langle e|D_q|g\rangle|^2~,
\end{align}
where $q$ denotes the standard component of the dipole associated to the scanning field polarization. Simulations of the magnetic sublevels energy and relative transition probabilities for $^{85}$Rb $F_g=2 \rightarrow F_e=3,4$ and  $^{87}$Rb $F_g=1 \rightarrow F_e=2,3$  transitions of $D_2$ line are known, and are based on calculation of the dependence of eigenvalues and eigenvectors of the Hamiltonian matrix on magnetic field for the whole hyperfine structure manifold \cite{Auzinsh_polarized_atoms,Tremblay_1990,Sargsyan_2014}.\\
\indent Magnetic field dependences of the frequency shifts and the relative probabilities of $\fbox{4} - \fbox{8}$ and 4 -- 8  transitions of $^{85}$Rb, as well as of $\textcircled{1} - \textcircled{3}$ and 1 -- 3 transitions of $^{87}$Rb for the case of $\sigma^+$ excitation are shown in Fig.~\ref{fig-shift_and_proba}. The $|g\rangle\rightarrow|e\rangle$ transition dipole moment for an atom interacting with longitudinal magnetic field is proportional \cite{Tremblay_1990} to
\begin{align}
|\langle e|D_q|g\rangle| \propto \sum_{F_e,F_g}c_{F_e'F_e}a(F_e,m_{F_e};F_g,m_{F_g};q)c_{F_g'F_g}~
\end{align}
with
\begin{equation}
\begin{split}
a(F_e,m_{F_e};&F_g,m_{F_g};q)=(-1)^{1+I+J_e+F_e+F_g-m_{Fe}}\\
&\times\sqrt{2J_e+1}\sqrt{2F_e+1}\sqrt{2F_g+1}\\
&\times\begin{pmatrix}
F_e & 1 & F_g \\
-m_{F_e} & q & m_{F_g}
\end{pmatrix}
\left\lbrace\begin{matrix}
F_e & 1 & F_g \\
J_g & I & J_e
\end{matrix}\right\rbrace,
\end{split}
\end{equation}
where the parentheses and the curly brackets denote the 3-$j$ and 6-$j$ coefficients, respectively. 

Good agreement between the theory and experiment is observed throughout the whole explored range of $B$-field (up to 1000~G), which proves complete consistency of the theoretical model. It should be noted that the calculations performed for other forbidden transitions of Rb $D_2$ line ($F_g = 3 \rightarrow F_e=1$ of $^{85}$Rb and $F_g=2\rightarrow F_e=0$ of $^{87}$Rb) show just a weak modification of the transition probabilities (several orders less than that for $F_g=2 \rightarrow F_e=4$ of $^{85}$Rb and $F_g=1 \rightarrow F_e=3$ of $^{87}$Rb), so that their experimental study is restricted by extremely small signal-to-noise ratio.\\
\indent The results of this study can be used for developing hardware and software solutions for wide-range optical magnetometers with nanometric (300 nm) local spatial resolution, as well as widely tunable frequency reference system, based on a NC and strong permanent magnets. Besides, formation of narrow (30 -- 50 MHz) optical resonances far away from the atomic transition (up to 20 GHz for strong $B$-field) allows realization of widely tunable laser frequency lock \cite{Muller_1998,Sargsyan_2014_OL}. Owing to small divergence of selective reflection radiation beam, which follows the divergence of incident radiation, the DSR signal can be easily detected at large distance ($\sim$~10 m) from the nanocell, which can be used for the remote optical monitoring of magnetic field. \\

\section{Conclusion}
Summarizing, we have studied, both experimentally and theoretically, the dynamics of frequency shifts and modification of probabilities of individual Zeeman components of Rb $D_2$ line hyperfine transitions in an external magnetic field up to 1000~G for $\sigma^+$ laser excitation, employing derivative selective reflection technique using $ ~L\sim~300$~nm vapor nanocell. It was shown, for the first time, that the probabilities of eight transitions, which are forbidden for $B$ = 0 (5 components for $^{85}$Rb $5S_{1/2}, F_g=2 \rightarrow 5P_{3/2}, F_e=4$ transitions and 3 components for $^{87}$Rb $5S_{1/2}, F_g=1 \rightarrow 5P_{3/2}, F_e=3$ transitions), undergo significant increase when applying magnetic field. Moreover, it was revealed that for the case of $\sigma^+$ excitation the intensity of $F_g=2, m_F=-3 \rightarrow F_e=4, m_F =-2$ transition of $^{85}$Rb becomes the largest among all the 25 Zeeman transitions of $F_g=2 \rightarrow F_e=1,2,3,4$ manifold in a wide range of magnetic field (200 -- 1000~G). The results of theoretical modeling, both for frequency shifts and modification of transition probabilities, are fully consistent with the experimental results.\\
\indent It is verified that the use of a half-wavelength-thick cell filled with Rb in DSR configuration is favorable for strong reduction of Doppler broadening of resonance lines (down to 60 MHz), allowing complete frequency resolution and quantitative study of each individual transition component.\\
\indent We should note that recent development of a new glass nanocell \cite{Whittaker_2015} can make the DSR technique a simple and reliable spectroscopic tool available for wider range of researchers and developers. The DSR method can be successfully employed also for studies of forbidden transitions of $D_2$ lines of Cs, K and Na, as well as for applications in optical magnetometry, metrology, and laser technologies.

\acknowledgments
The authors are grateful to A. Sarkisyan for his valuable contribution in fabrication of the NC. C.L. is grateful to Prof. Oleg Ulenikov for valuable discussions. Research conducted in the scope of the International Associated Laboratory IRMAS (CNRS-France \& SCS-Armenia)

\end{document}